\documentclass[12pt]{iopart}

\usepackage[dvips]{graphicx}   
\newcommand{\beqn}{\begin{equation}}
\newcommand{\eeqn}{\end{equation}}
\newcommand{\beqna}{\begin{eqnarray}}
\newcommand{\eeqna}{\end{eqnarray}}

\begin{document}

\title[Long-range memory and multifractality in gold market]{Long-range memory and multifractality in gold markets}

\author{Provash Mali, Amitabha Mukhopadhyay}

\address{Physics Department, North Bengal University, Darjeeling 734013, West Bengal, India}
\ead{provashmali@gmail.com}
\begin{abstract}
Long-range correlation and fluctuation in the gold market time series of world's two leading gold consuming countries, namely China and India, are studied. For both the market series during the period 1985-2013 we observe a long-range persistence of memory in the sequences of maxima (minima) of returns in successive time windows of fixed length, but the series as a whole are found to be uncorrelated. Multifractal analysis for these series as well as for the sequences of maxima (minima) is carried out in terms of the multifractal detrended fluctuation analysis (MF-DFA) method. We observe a weak multifractal structure for the original series that is mainly originated from the fat-tailed probability distribution function of the values, and the multifractal nature of the original time series is enriched into their sequences of maximal (minimal) returns. A quantitative measure of multifractality is provided by using a set of ``complexity parameters''.      
\end{abstract}
\section{Introduction}
Gold is a universally accepted precious metal and it is an extremely popular investment instrument. Gold is also considered as a kind of hard currency that has a very high degree of market liquidity. The day-to-day fluctuations of gold market rate seems to be quite interesting, and even to the regular traders the fluctuation pattern is so random that often it becomes almost impossible to predict its accurate rise or fall. There are several factors that (in)directly influence the gold market e.g., (i) the rise and fall of U.S. currency value that usually follows an inverse relation with the gold market, (ii) the geopolitical situation, (iii) the local/global financial crisis, (iv) the supply-demand mismatch and (v) the inflation rate etc.. With so many parameters the market dynamics of gold is always quite complex and to understand the same it requires a rigorous study from all possible perspectives. In recent years the so-called MultiFractal Detrended Fluctuation Analysis (MF-DFA) \cite{Kant02} is found to be one of the highly successful tools for characterizing the nonstationary time series. The MF-DFA technique has so far been applied in various fields of stochastic systems e.g., in the stock market analysis \cite{Matia03,Oswi05,Kwap05,Oswi06}, in geophysics \cite{Kant03,Movahed06,Eva06,Mali14t}, in biophysics \cite{Esen11,Kumar13,Card12} and also in various branches of basic and applied physics \cite{Zhang08,Mali14c,Igna10}. Obviously the list of references on the applicability of the MF-DFA methodology given here is not a complete one. In this work we employ the MF-DFA technique to analyze the gold market time series of world's two leading gold consuming countries namely China and India for the period 1985-2013. It is to be noted that the combined gold consumption of these two countries is about $50\%$ of the global demand \cite{WGC}. So we choose these two markets only for a case study, otherwise there is no other intention added to the choice. China at present is the number one producer and consumer of gold. On the other hand though India is not a significant gold producing country, like China it is also one of the largest gold consuming nations, most of which is imported. The gold markets in both countries however, are not yet fully liberalized, and expectedly one component of the market fluctuations is the economic and financial policies adopted by the respective government. About $84\%$ of the total gold supplied to mainland China's gold market goes to manufacturing jewellery, bar, coin etc. and to the technology. At the end of 2013 about $60\%$ of all private sector gold demand in China (1066 tonnes) went only to jewellery, while at the same time at US\$ $77$bn the investors' gold holdings are small in comparison to other available avenues of investment. At the same period of time the private sector gold demand in India stood at 975 tonnes, about $80\%$ of which went to jewellery manufacturing, about $15\%$ to investment purposes and about $5\%$ to various industries \cite{WGC}.

In some recent reports on gold market analysis \cite{Bolgorian11,Wang11,Ghosh12} it has been observed that the gold return time series are multifractal in nature, the main source of which is a long-range temporal correlation present in the data. But it is obvious that the price returns or any other time series off-shoot cannot be persistently correlated over some reasonable time duration to the level where it can be exploited to gain excess returns \cite{Odean99,Liu99}. Otherwise, this fact could easily be exploited for price movement prediction that would ultimately eliminate the correlations. Hence one cannot expect to have easily exploitable correlation patterns in financial time series and also in the market values of a highly liquid commodity like gold. However, this certainly does not mean that correlations do not exist at all. 
Our recent observation \cite{Mali14a,Mali14b} shows that the multifractality in the gold return time series is originated not only from a long-range correlation but also from a fat-tailed probability function of the values. The observation in \cite{Mali14a,Mali14b} apparently contradicts those obtained from some other previous analyses \cite{Bolgorian11,Wang11,Ghosh12}. Moreover, our autocorrelation analysis (to be discussed in section 3.1) shows that the gold return time series do not possess autocorrelation. Therefore, one would expect a hidden source of long-range correlation in the gold market returns. In fact, one of the most mysterious properties of any market series is the presence of long-term correlation patterns in the variance of returns known as the volatility clustering \cite{Admati88,Wang08}. A quantitative manifestation of it is that, while returns themselves are uncorrelated, the absolute return $|R_t|$ (also called the `volatility') or its square shows an autocorrelation that follows a power-law type of scaling relation within a moderate range of time lag $\tau$ \cite{Mandelbrot63}. In \cite{Muchnik09} it is shown that a particular sequence of return record may also exhibit long-range correlation that may add further insight into the market dynamics. Here we show that, there are some (sub)sequences at least in the Chinese and Indian gold market return series that exhibit a moderate amount of long-term persistence in a way that can be compared with a correlation exponent ($\gamma$), and the persistence effect introduces a significant amount of multifractality in these markets. The sequences taken here are the maxima (minima) of the returns in successive time windows of fixed length $R$ (= 5 and 10 trading days). Note that if a sequence of maximal (minimal) returns is mostly of the same sign, as it is seen here, the sequence may be taken as a proxy to the market volatility.      
The rest of the article is arranged as follows. In section 2 we describe the data used in this analysis. The methodology and the results are presented in section 3, where under two different subsections we describe the autocorrelation analysis and the detrended analysis. The paper is summarized in section 4. 
\section{Data}
As mentioned above the gold markets in both the countries are not completely free and they are substantively influenced by the policies taken by the respective government. In 1950 People's Republic of China put its gold market under the strict control of the state. Private use of precious metals was banned, while all export and/or import activities in this regard used to be controlled by the state via the People's Bank of China (PBoC). An economic reform started in 1978 and a gold market was established under the close control of PBoC, a monopoly that lasted till 2001. Until 2004 bullion investment was effectively prohibited in China. Since 2001 with the establishment of Shanghai Gold Exchange, the state control is gradually reduced. However, China has still to go a long way to completely free its gold market from the state control. One major step in this regard may be to stop undervaluing the local currency, the Chinese Yuan. As the political and administrative systems prevailing in the country are controlled by a single party, certainly China is moving towards a free market with a steady and rapid pace. In contrast historically there has always been less control of the state on the Indian gold market. People from time immemorial can use, buy and sell precious metals. The economic reforms in the country started in 1991. However, in India the progress in this regard has been rather slow. The Indian gold market to a large extent depends on the import, and the government still continues to take measures like hiking import duties, restrict import quotas etc., to control the supply of this precious metal, which has a strong and direct bearing on the gold market. Mainly because of its multi-party political system and a functional democracy, India does not afford the luxury of a unanimous view on most of the economic issues, and therefore, it cannot move very fast in the path of liberalization.
\begin{figure}
\centering
\includegraphics[width=6in,height=4.5in]{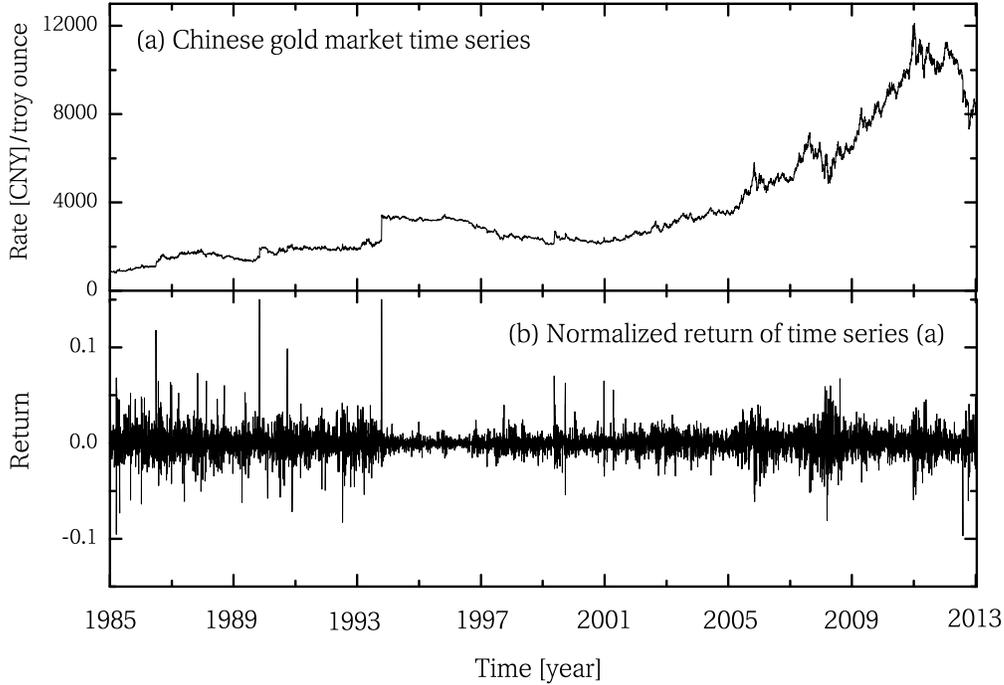}
\caption{(a) The gold market time series in China for the period 1985-2013 and (b) the corresponding returns.}
\label{Fig:Rate}
\end{figure}
\begin{figure}
\centering
\includegraphics[width=5in,height=4in]{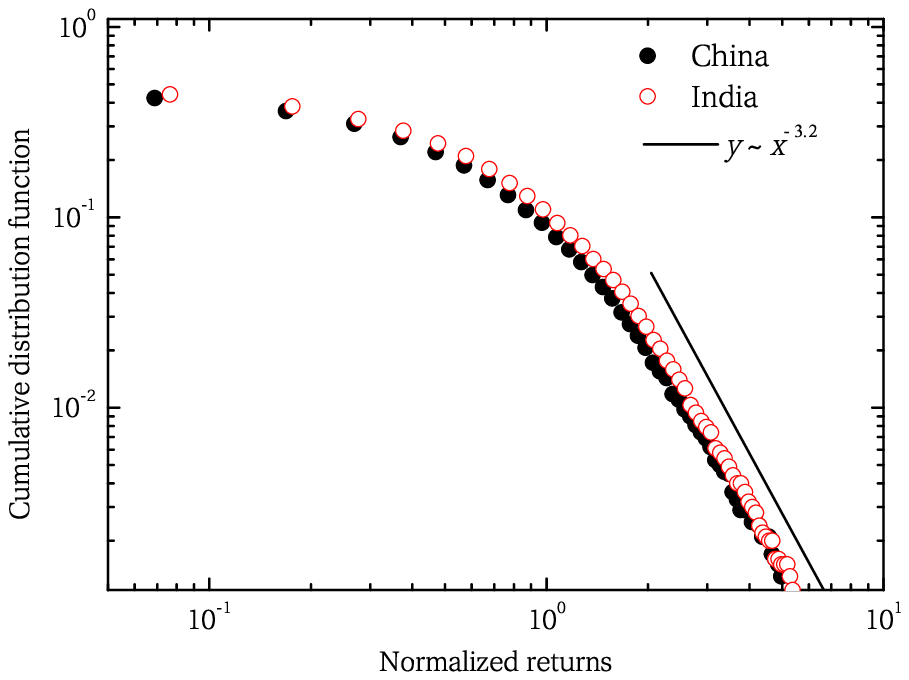}
\caption{The cumulative distribution functions of logarithmic gold returns for the Chinese (black filled circles) and Indian (red empty circles) market returns.}
\label{Fig:cdf}
\end{figure}

The gold price time series data used here are taken from the database of the World Gold Council \cite{WGC}. The logarithmic difference between two successive trading days, also known as return, is calculated as: $R_t = \ln P_{t+1} - \ln P_t$, where $P_t$ in the closing price at day $t$. This definition can trivially be generalized to returns computed for any arbitrary time lag $\tau$. We show in Fig. \ref{Fig:Rate} the market trend of China for the period 1985--2013, where diagram (a) represents the original time series, whereas (b) shows the corresponding return series. The figure shows how the gold market of China has evolved with time over the last 28 years. The Indian gold market also follows more or less a similar pattern, and hence the corresponding series is not graphically shown. These price series are measured in local currencies, i.e., the Cinese in Cinese Yuan and the Indian Indian Rupees. To check to what extent the relative currency fluctuation affects the currency denominated time series, we have converted the Rupee denominated Indian time series to one denominated by the Chinese Yuan, taking the day to day currency conversion rate \cite{OANDA}. However, this is done for a limited period of time (1994--2013) and not for the entire time period (1985--2013) of the original analysis. The consequences of currency conversion are, as we shall see later, substantive. The cumulative distribution functions (CDFs) of the normalized returns $r_t$ for the Chinese and the Indian market series are shown in Fig. \ref{Fig:cdf}. We define $r_t$ as
 \beqn 
 r_t = \frac{R_t - \left< R_t \right>_T}{\sigma_T},
 \eeqn
where $\left< ~\right>_T$ indicates the mean of $R_t$ over the considered time period $T$, and $\sigma_T$ is the standard deviation of returns over $T$.
We find the tail exponent $\zeta \approx 3.2$ i.e., each CDF follows almost an inverse cubic power-law with normalized return, and hence the underlying probability distribution may be considered as a fat-tailed function. The first indication of such a power-law type of CDF of market return data can be traced back to \cite{Lux96}. For the stock markets \cite{Gopi98,Gopi99} as well as for the gold markets during the period 1968-2010 \cite{Bolgorian11}, the CDFs were also found to be inverse cubic.
\section{Analysis and Results}
\subsection{Autocorrelation function}
Consider a time series $\{x_i:i=1,2,\cdots,N\}$ where the index $i$ corresponds to the time of measurement. Autocorrelation function provides a correlation between the $i$th and the $(i+s)$th measurement for different values of time lag $(s)$. In order to remove a constant offset in the series, the mean of the series $\left<x\right>=(1/N)\sum_{i=1}^N x_i$ is usually subtracted and a new varibale $\bar{x}_i=x_i-\left<x\right>$ is introduced. Then the auto-covariance between any two $\bar{x}$s separated by $s$ steps is defined as
\begin{equation}
C'(s) = \left<\bar{x}_i \bar{x}_{i+s}\right> = \frac{1}{N-s}\sum_{i=1}^{N-s}\bar{x}_i \bar{x}_{i+s}. 
\end{equation}
When the above $C'(s)$-function is normalized by the variance $\left<\bar{x}^2_i\right>$, the function is called an autocorrelation function $C(s)$. If $\{x_i\}$-s are uncorrelated then for any $s>0$, $C(s)=0$. The series has short-range correlation if $C(s)$ declines exponentially i.e., $C(s) \propto \exp(-s/s_0)$ with some characteristic $s_0$. On the other hand for a long-range correlated series $C(s)$ declines as a power-law like, $C(s) \propto s^{-\gamma}$ with the exponent $0 < \gamma <1$. Due to the noise superimposed on the data $x_i$ and due to the underlying trends of some unknown origin, a direct measurement of $C(s)$ is usually not possible. Hence, the exponent $\gamma$ is estimated indirectly. Here we employ the MF-DFA method in order to capture the nature of correlation (if any) present in the analyzed time series data. However, as a preliminary estimate of $\gamma$ we study the autocorrelation functions. Note that a stationary fluctuating series can also be characterized by the so-called power spectrum $E(f)$ with frequency $f$ as, $E(f) \sim f^{\beta}$. However, for a stationary time series the exponent $\beta = 1 - \gamma$.
\begin{figure}[t]
\centering
\includegraphics[width=6.5in,height=5.5in]{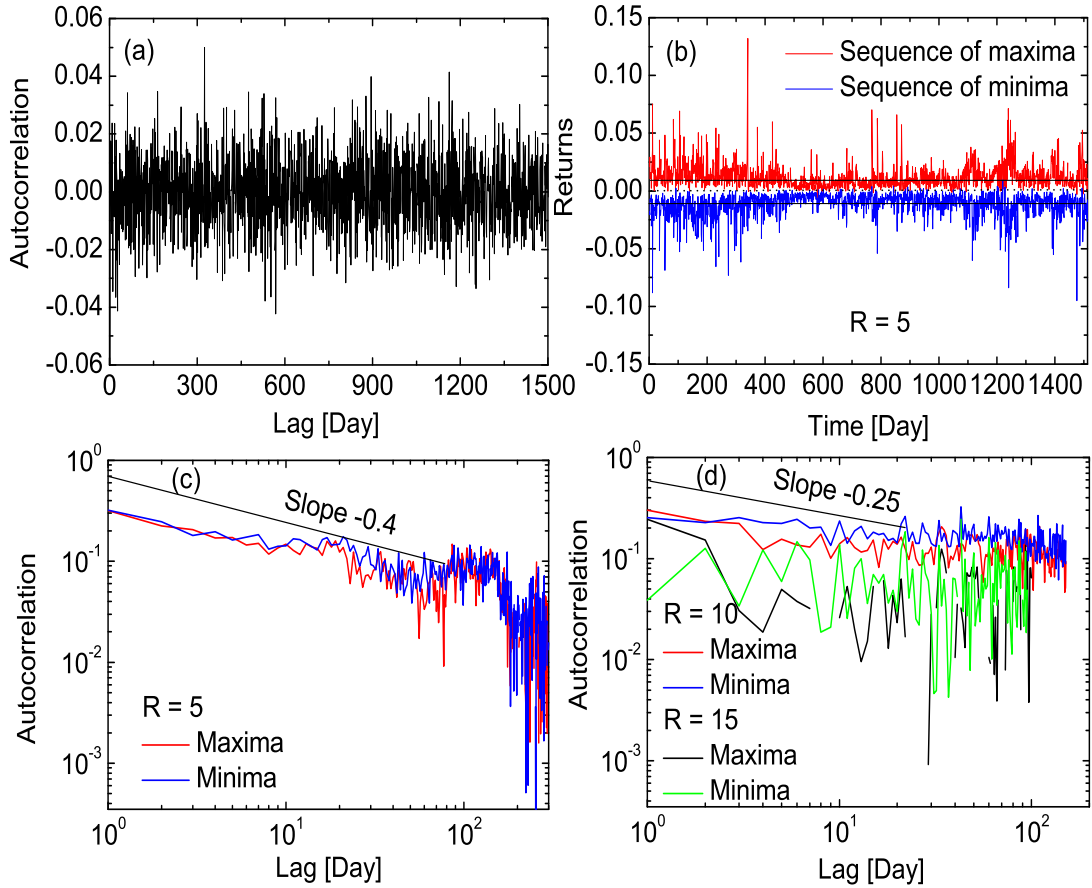}
\caption{(a) Autocorrelation function for the gold market return for the Indian market series. (b) Sequences of maximal (red) and minimal (blue) returns in successive time windows of length $R=5$ (trading days) for the Indian market. (c) Autocorrelation functions corresponding to the maximal and minimal returns of (b).  (d) The same as (c) but for $R=10$ (red and blue) and for $R=15$ (black and green). Notice that the total time series do not show any autocorrelation but the sequences of maxima(minima) for $R=$ 5 and 10 show autocorrelation functions that decline following a power-law (signature of long-range correlation). The autocorrelation disappears for $R=15$.}
\label{Fig:Corr}
\end{figure}

In Fig. \ref{Fig:Corr} we illustrate our results on autocorrelation analysis, where only the results of Indian gold market time series are shown. More or less similar results are also obtained for the Chinese market and therefore, the corresponding plots are not shown. We do not observe any correlation in the gold market returns for the world's leading gold consuming country (China), and the same is as well true for the Indian gold market [Fig. \ref{Fig:Corr}(a)]. Within errors the autocorrelation function $C(s)$ vanishes for all $s$. As mentioned the long-range correlation may also originate from some sequences present in the series. For example, we consider the sequence of maximal (minimal) values of returns in successive time windows of fixed length $R$. Note that the sequences are constructed out of the original series of daily logarithmic returns by selecting the maxima (minima) of daily returns over each consecutive intervals of fixed length $R$. Obviously both the sequences for a fixed $R$ contain $R$ times fewer points than the original one. Figure \ref{Fig:Corr}(b) shows the sequences of maxima (red) and minima (blue) of returns for $R=5$ obtained from the Indian gold market series. In this diagram one can see pronounced patches of small and large maxima and minima (above and below the mean) that are clumped together. The patches demonstrate qualitatively the occurrence of a memory effect (clustering), where large maxima (minima) tend to follow large maxima (minima) and small maxima (minima) tend to follow the small ones. We calculate the autocorrelation functions for the sequences for three different values, $R = 5,~10,~\mbox{and}~15$ trading days. Figure \ref{Fig:Corr}(c) and (d) illustrate the results of this calculation. Figure \ref{Fig:Corr}(c) with $R=5$ shows the presence of long-range correlation in both the sequences that results in an autocorrelation function that follows a power-law type of scaling relation with time lag. A signature of long-range correlation in the sequences with $R=10$ is also visible [Fig. \ref{Fig:Corr}(d)], but the correlation is washed out at window length $R = 15$. Apparently with $R=15$ the subseries become statistically too weak to show any correlation over a reasonable value of the time lag, and for $R=15$ the statistical noise also becomes significant. These could be the reasons of an instability in $C(s)$ at $R=15$. In our subsequent analysis we shall consider the sequences of maximal and minimal returns only for $R=5~ \mbox{and}~ 10$.
\subsection{Multifractal detrended fluctuation analysis}  
Now-a-days the MF-DFA formalism \cite{Kant02} has become a standard tool for the time series data analysis. Without claiming any originality a brief description of the methodology is outlined below. Let $\{x_k:~k=1,~2, \dots,N\}$ be a time series of length $N$. The MF-DFA technique consists of five following steps:
\begin{enumerate}
\item Determine the profile
\beqn
Y(i) = \sum_{k=1}^{i} [x_k - \left<x\right>],~~ i=1,~2,\dots, N,
\label{eq:Profile}
\eeqn
where $\left<x\right> = (1/N)\sum_{k=1}^N x_k$ is the mean value of the analyzed time series.
\item  Divide the profile $Y(i)$ into $N_s = int(N/s)$ non-overlapping segments of equal length $s$. Depending upon the length of the series one has to choose an appropriate $s$ value. In case the length $N$ is not a multiple of the considered time scale $s$, the same dividing procedure is repeated starting from the opposite end of the series. Hence in order not to disregard any part of the series usually altogether $2N_s$ segments of equal length are obtained.
\item Calculate the local trend for each of the $2N_s$ segments. This is done by a least-square fit to the data present in individual segments. Linear, quadratic, cubic or even higher order polynomials may be used to detrend the series, and accordingly the procedure is said to be the MF-DFA1, MF-DFA2, MF-DFA3, $\dots$ analysis. Let $y_p$ be the best fitted polynomial to an arbitrary segment $p$ of the series. Then determine the variance
\beqn
F^2(p,s) = \frac{1}{s}\sum_{i=1}^{s}\left\{Y[(p -1)s+i] - y_{p}(i) \right\}^2
\eeqn
for $p=1, \dots N_s$, and for $p = N_s+1, \dots, 2N_s$ it is given as,
\beqn
F^2(p,s) = \frac{1}{s}\sum_{i=1}^{s}\left\{Y[N-(p -N_s)s+i] - y_{p}(i) \right\}^2.
\eeqn
\item Define the $q$th order MF-DFA fluctuation function
\beqn
F_q(s) = \left\{ \frac{1}{2N_s} \sum_{p=1}^{2N_s}[F^2(p,s)]^{q/2} \right\}^{1/q}
\label{eq:Fq1}
\eeqn
for all $q \neq 0$ and for $q=0$ it is given as
\beqn
F_q(s) = \exp\left\{ \frac{1}{4N_s} \sum_{p=1}^{2N_s}\ln [F^2(p,s)] \right\}.
\label{eq:Fq2}
\eeqn
\item Then the scaling behavior of the fluctuation functions is examined for several different values of the exponent $q$. If the series $\{x_k\}$ possesses long-range (power-law) correlation, $F_q(s)$ for large $s$ would follow a power-law type of scaling relation like
\beqn
F_q(s) \sim s^{h(q)}.
\label{eq:Scaling}
\eeqn
\end{enumerate}
In general the exponent $h(q)$ depends on $q$ and is known as the generalized Hurst exponent. The exponent $h(2)$ is related to the correlation exponent $\gamma$ and the power-spectrum exponent $\beta$ by
\beqn
h(2) = 1 - \gamma/2 = (1+\beta)/2.
\eeqn
For a stationary time series $h(2) =H$ -- the well known Hurst exponent \cite{Kant03}. On the other hand for a monofractal series $h(q)$ is independent of $q$, since the variance $F^2(p,s)$ is identical for all the subseries and hence Eqns. (\ref{eq:Fq1}) and (\ref{eq:Fq2}) yield identical values for all $q$. Note that the fluctuation function $F_q(s)$ can be defined only for $s \geq m+2$, where $m$ is the order of the detrending polynomial. Moreover, $F_q(s)$ is statistically unstable for very large $s~(\geq N/4)$. If small and large fluctuations scale differently, there will be a significant dependence of $h(q)$ on $q$. For positive / negative values of $q$, $F_q(s)$ will be dominated by large / small variances which correspond to the large / small deviations from the detrending polynomial. Thus for positive / negative values of $q$, $h(q)$ describes the scaling behavior of the segment with large / small fluctuations.
\subsubsection{Relation with standard multifractal variables:}
One can easily relate the $h(q)$ exponent with the standard multifractal exponent, such as the multifractal (mass) exponent $\tau(q)$. Consider that the series $\{x_k\}$ is a stationary and normalized one. Then the detrending procedure (step 3) of the MF-DFA methodology is not required, and the variance of such series $F^2_N$ is given by
\beqn
F^2_N(p,s) = \{Y(p s) - Y[(p-1)s]\}^2.
\label{eq:Vari2}
\eeqn
Under such circumstanes the fluctuation function and its scaling law are given by
\beqn
F_q(s) = \left\{\frac{1}{2N_s} \sum_{p = 1}^{2N_s}|Y(p s) - Y[(p-1)s]|^q \right\}^{1/q} \sim s^{h(q)}.
\label{eq:Fq2x}
\eeqn
Now if we assume that the length of the series $N$ is an integer multiple of the scale $s$, then the above relation can be rewritten as,
\beqn
\sum_{p=1}^{N/s} |Y(p s) - Y[(p -1)s]|^q \sim s^{qh(q) -1}.
\label{eq:Fq3x}
\eeqn
In the above relation the term under $|\cdot|$ is nothing but the sum of $\{x_k\}$ within an arbitrary $p$th segment of length $s$. In the standard theory of multifractals it is known as the box probability $\mathcal P(s,p)$ for the series $x_k$. Hence,
\beqn
\mathcal P(p,s) \equiv \sum_{k=(p-1)s+1}^{p s} x_k = Y(p s) - Y((p -1)s).
\label{eq:BoxPro}
\eeqn
The multifractal scaling exponent $\tau(q)$ is defined via the partition function $Z_p(s)$
\beqn
Z_{\mathcal P}(s) \equiv \sum_{p=1}^{N/s} |\mathcal P(p,s)|^q \sim s^{\tau(q)},
\label{eq:Z}
\eeqn
where $q$ is a real parameter.
From Eqns.~(\ref{eq:Fq3x})--(\ref{eq:Z}) it is clear that the multifractal exponent $\tau(q)$ is related to $h(q)$ through the following relation:
\beqn
\tau(q) = q\,h(q) - 1.
\label{eq:tau}
\eeqn
Knowing $\tau(q)$ one can calculate the most important parameter of a multifractal analysis -- the multifractal singularity spectrum (also called the spectral function) $f(\alpha)$, which is related to $\tau(q)$ through a Legendre transformation \cite{Feder88}: $\alpha = \partial \tau(q)/\partial q$. The $f(\alpha)$-function is defined as
\beqn
f(\alpha) = q \alpha - \tau(q)
\label{eq:fa}
\eeqn
Here $\alpha$ is the singularity strength or H\"{o}lder exponent. The singularity spectrum gives a parametric representation of a fractal analysis. In other words it is a measure of fractal dimension. For a monofractal signal it turns out to be a delta function at the corresponding $\alpha$, whereas for a multifractal signal $f(\alpha)$ produces a typical concave downward parabolic spectrum and the degree of multifractality is characterized by a wide aperture and relatively small asymmetry of $f(\alpha)$.  
\subsection{Results of MF-DFA analysis}
\begin{figure}[t]
\centering
\includegraphics[width=6.5in,height=5in]{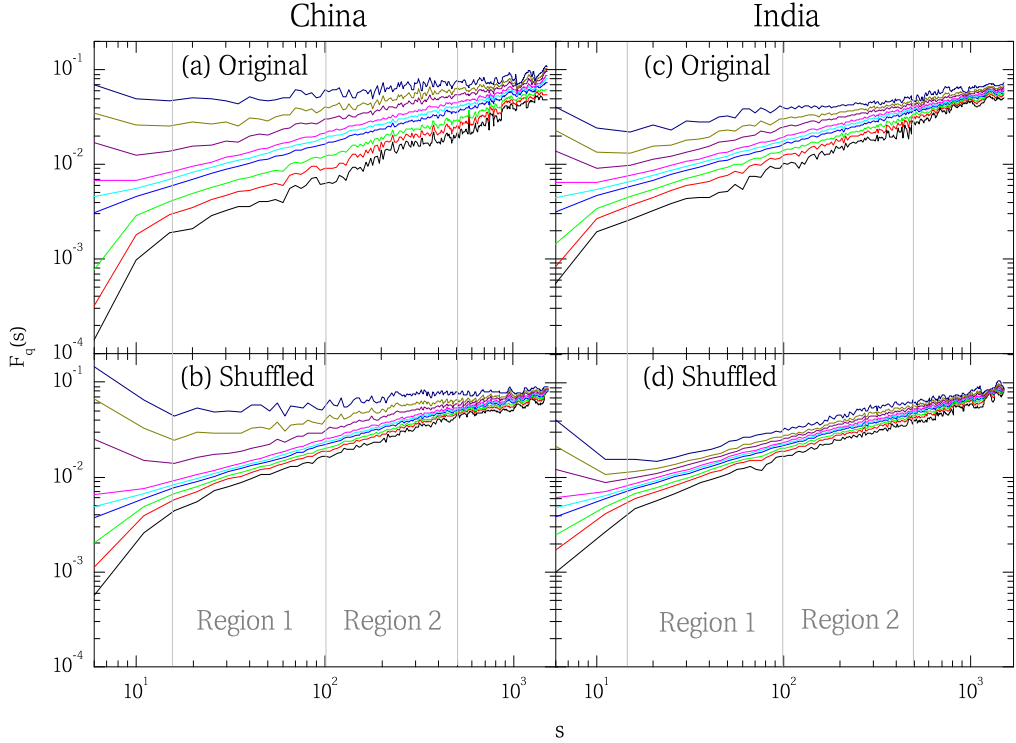}
\vspace{-2cm}
\caption{The MF-DFA2 fluctuation functions for the gold market returns in China (left panel) and India (right panel). The lower panel shows the fluctuation functions generated from the shuffled series corresponding to the original ones. In all the cases starting from the bottom to the top the curves represent $q=-10$, -5, -3, -1, 0, 1, 3, 5 and 10, respectively.} 
\label{Fig:Fq1}
\end{figure}

We calculate the MF-DFA fluctuation functions $F_q(s)$ for the Chinese and the Indian gold market returns as well as for the corresponding sequences of maxima (minima) for two different values of time window  $R=5$ and 10 (days). As a second order polynomial is used for the detrending purpose (as mentioned in step (iii) of the methodology), the $F_q(s)$-functions are said to be the MF-DFA2 fluctuation functions. The exponent $q$ is varied from $-10$ to $+10$ in steps of $0.5$ and the scale parameter (time) $s$ is varied from $6$ to $N/5$, where $N$ is the length of the series. In Fig.~\ref{Fig:Fq1} we show the scaling behavior of some of the $F_q$-functions calculated for the gold market returns of China (left) and India (right). We also show in the lower panel of the figures the $F_q$-functions estimated from a series shuffled at random that correspond to the respective original time series.
\begin{figure}
\centering
\includegraphics[width=6.5in,height=5.5in]{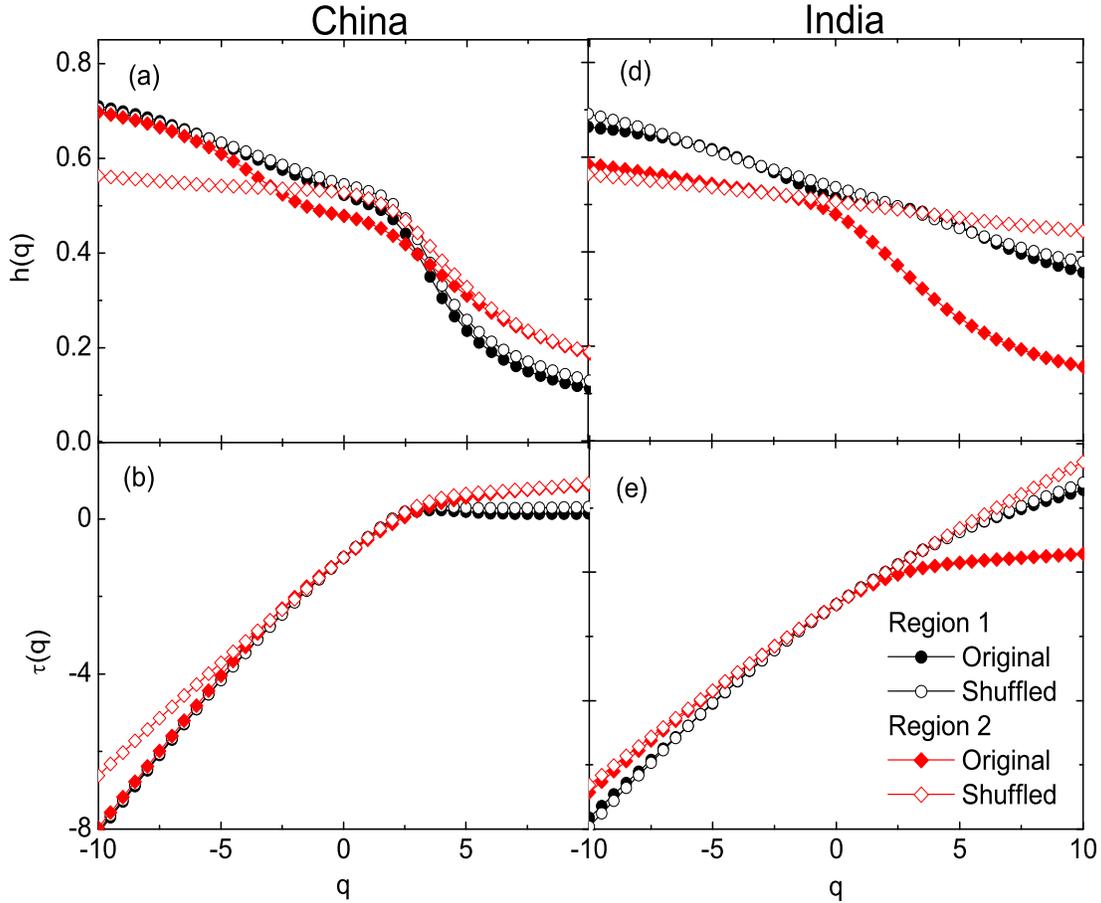}
\vspace{-1cm}
\caption{(a) The generalized Hurst exponent spectra (upper panel) and the multifractal exponent spectra (lower panel) for the analyzed gold market returns. The spectra are calculated in two different regions of the $\ln F_q(s) $ vs. $\ln s$ data (see Fig. \ref{Fig:Fq1}).} 
\label{Fig:Hq1}
\end{figure}
\begin{figure}[t]
\centering
\includegraphics[width=6.5in,height=7.5in]{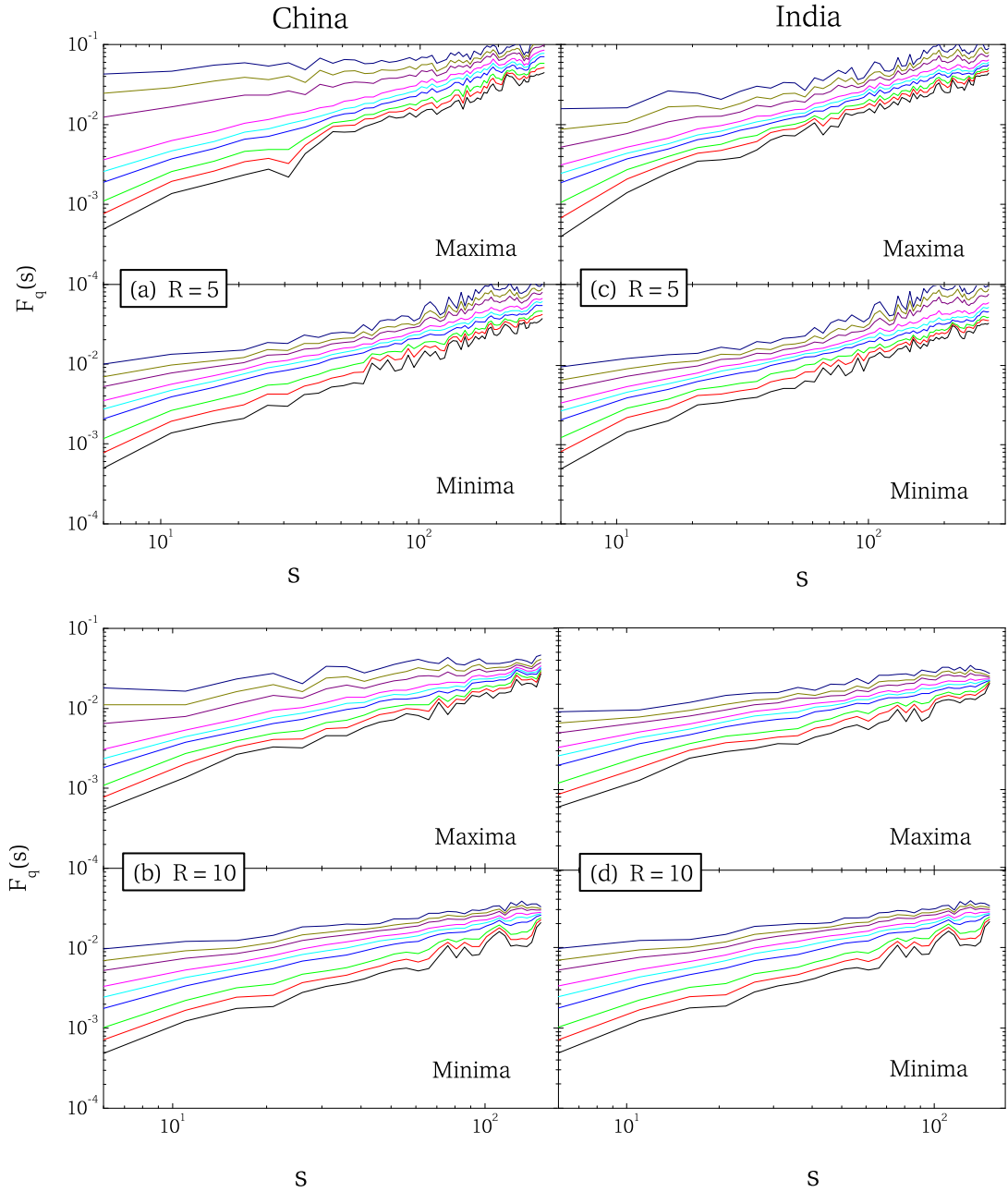}
\vspace{-1cm}
\caption{The MF-DFA2 fluctuation functions for the sequences of maxima(minima) for $R=5$ (upper) and 10 (lower). The left (right) panel represents the market series of China (India). In all the cases starting from the bottom to the top curves represent $q=-10$, -5, -3, -1, 0, 1, 3, 5 and 10, respectively.} 
\label{Fig:Fq2}
\end{figure}
\begin{figure}
\centering
\includegraphics[width=6.5in,height=5in]{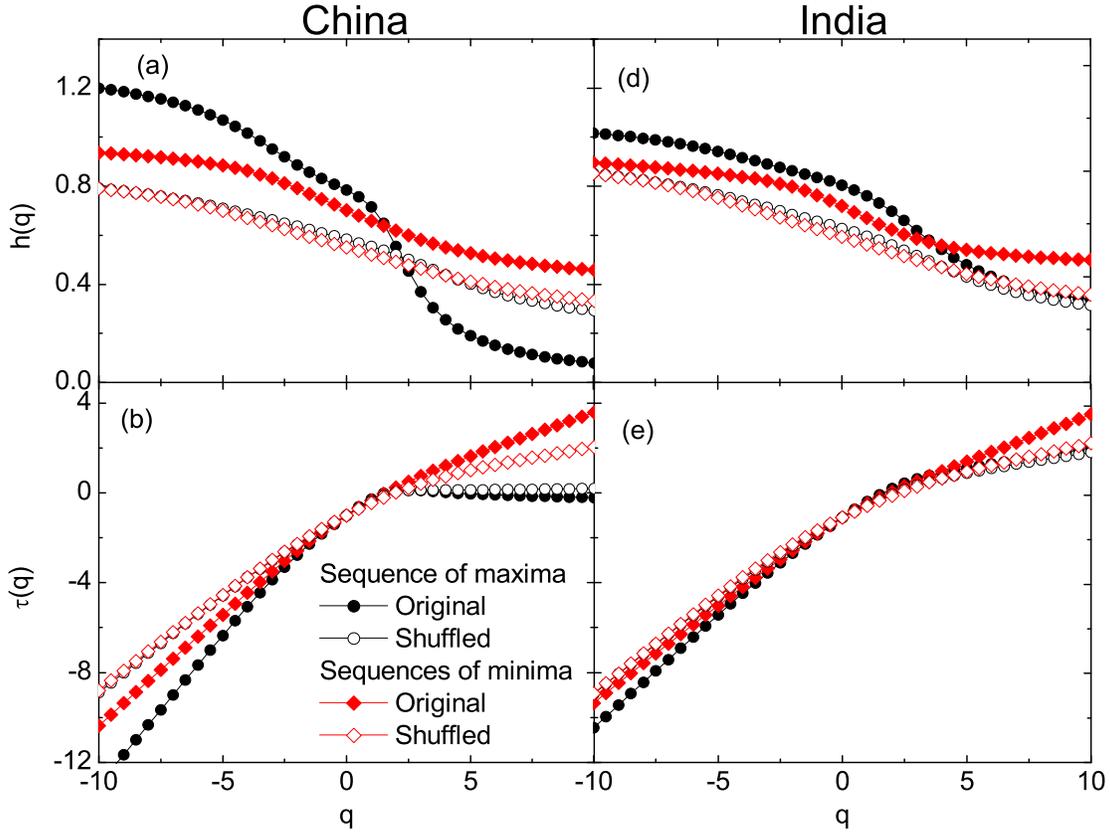}
\caption{The generalized Hurst exponent spectra (upper panel) and the multifractal exponent spectra (lower panel) obtained from the sequences of maxima (black) and minima (red) for $R=5$. Left (right) panel represents the results from the Chinese (Indian) market series. Predictions of the shuffled series corresponding to each of the original series are also shown.} 
\label{Fig:Hq5}
\end{figure}
\begin{figure}
\centering
\includegraphics[width=6.5in,height=5in]{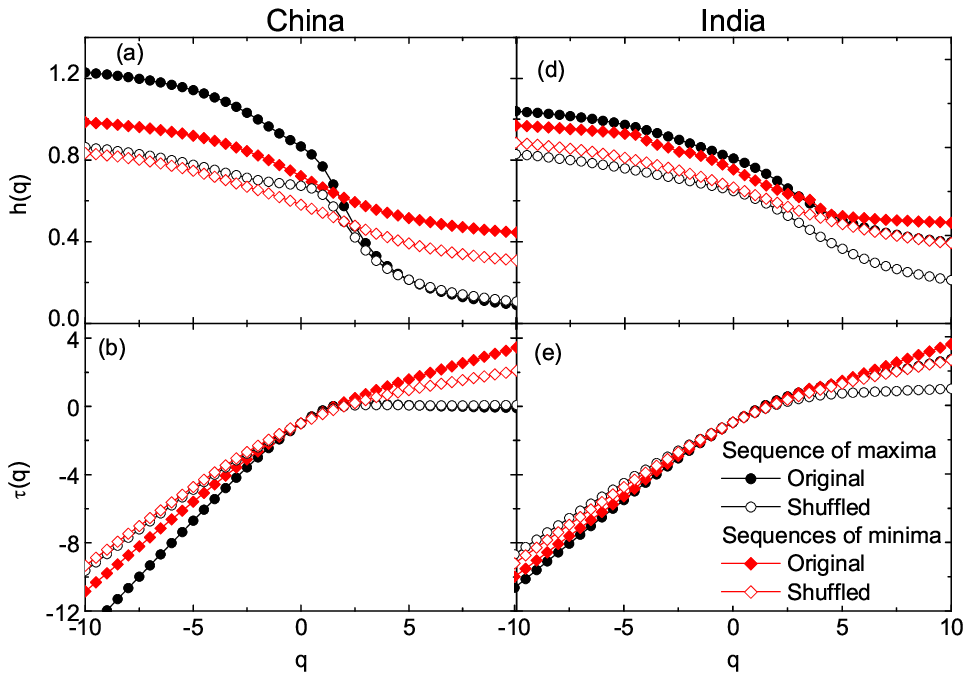}
\caption{The same as Fig. \ref{Fig:Hq5} but for $R =10$.} 
\label{Fig:Hq10}
\end{figure}

The importance of analyzing a shuffled series for a given time series data is that, a direct comparison between them gives an insight into the origin(s) of multifractality present (if any) in the time series data. It is a known fact that there might be two different sources of multifractality present in a time series data, namely (i) multifractality due to long-range temporal correlations of the small and large fluctuations and (ii) multifractality due to a fat-tailed probability distribution function of the values. Multifractality of the first kind can be removed by a random shuffling of the given series and the resultant shuffled series will exhibit monofractal scaling. Multifractality of the second kind will remain intact even after the shuffling, since the probability distribution will not alter by a random shuffling. If a given time series consists of both kinds of multifractality, the corresponding shuffled series will exhibit weaker multifractality than the actual series.

It is clear from Fig.~\ref{Fig:Fq1} that the scaling behavior of $F_q(s)$ for both the gold market time series as well as for their shuffled counterparts more or less follow the same type of scaling-law Eqn. (\ref{eq:Scaling}), but within a limited scale interval. Note that in the $s < 15$ regions of the $F_q(s)$ values suddenly start deviating form a smooth behavior (\ref{eq:Scaling}), while from $s \geq 15$ the $F_q(s)$ values for different $q$ start to converge. Notable fluctuations can be seen at large scale ($s \ge 500$). In fractal theory a power law type scaling behavior is expected only at large $s$. This may be the reason of sudden deviation in $F_q(s)$ at small $s$. Moreover a close scrutiny of the $F_q(s)$ functions indicates that a single $h(q)$ exponent cannot scale the entire $15 \leq s \leq 500$ interval [see Fig.~\ref{Fig:Fq1}(c)] over which the $F_q(s)$ functions behave systematically. There exist noticeable changes in the slope at about $s = 100$. Therefore, to extract $h(q)$ we choose two different scale intervals of the $F_q(s)$ versus $s$ plots--region I ($15 \leq s \leq 100$) and region II ($100 \leq s \leq 500$), as shown in Fig. \ref{Fig:Fq1}. The reason behind two different strengths of the scaling exponent is not very much clear at this point.

The $h(q)$ spectra calculated in two different scale ($s$) intervals for both the market series as well as for their shuffled values are shown in Fig. \ref{Fig:Hq1}. The $\tau(q)=q h(q)-1$ spectra are given in the lower panel of the same figure. Apparently the $q$ dependence of the $h(q)$ spectra and the nonlinearity of the $\tau(q)$ spectra imply the existence of multifractality in the analyzed gold markets. However, the exponent $h(2) = 1-\gamma/2$ for all the series is very close to or even less than $0.5$ that results the correlation exponent $\gamma \geq 1$. This is consistent with our autocorrelation analysis [Fig. \ref{Fig:Corr}(a)] i.e., there is no direct signature of correlation present in the data. Moreover, the differences in the results obtained from the original series and the corresponding shuffled series are insignificant. Thus, we speculate that the source(s) of multifractality in the gold market time series, at least for the Chinese and Indian markets, might be (i) a fat-tailed probability distribution function (PDF), or (ii) a short / long-range temporal correlation in the sequences of maximal (minimal) returns over successive non-overlapping time windows, or (iii) even both of the said factors. This observation in another sense is our motivation behind this work. The difference between the original and the shuffled series generated $h(q)$ values in region II is slightly greater than that in region I. This is an indication of the presence of a weak long-range correlation in the data that may be present in the sequences of maxima (minima). The reason of almost null difference between the original and shuffled series generated values of $h(q)$ (or $\tau(q)$) in region I may be that, for small scale ($s$) the profile series is divided into a large number of small segments, which may not be able to capture the information on long-range correlation. The observed multifractality in this cases is mainly because of the fat-tailed probability density function. A quantitative description on the source(s) of multifractality is given at the end of this section.

The MF-DFA2 fluctuation functions for the sequences of maxima (minima) for $R=5$ and 10 are shown Fig. \ref{Fig:Fq2}. These $F_q$-functions are also found to obey the scaling relation (\ref{eq:Scaling}) well, but within the limited scale interval $s \approx 5 - 75$. Once again we have extracted the $h(q)$ exponents from linear regressions to the log-log data of $F_q(s)$ versus $q (\leq 75)$. The $h(q)$-spectra along with the corresponding $\tau(q)$ spectra for $R = 5$ are shown in Fig. \ref{Fig:Hq5}, where left (right) panel is drawn for the Chinese (Indian) market. The similar results for $R=10$ are shown in Fig. \ref{Fig:Hq10}. In these figures also we compare the predictions of the original series (actually, the sequences of maximal/minimal returns) to that of their shuffled ones. Stronger order dependent $h(q)$ spectra are obtained in the case of the sequences of maxima than the sequences of minima. It is seen that the order dependence of the generalized Hurst exponents calculated from the shuffled series is much weaker than the corresponding empirical values. The $h(2)$ exponent values are all within the limits: $0.5 \leq h(2) \leq 1.0$. All these observations are signatures of long-range correlations present in the sequences of maximal (minimal) gold returns. Obviously, the nonlinearity in the $\tau(q)$ spectra follows from the $h(q)$ spectra and therefore, it does not require further discussion. Our observations for the sequences with $R=10$ and with $R=5$ are more or less similar.

The most important observable of a multifractal analysis is the multifractal singularity spectrum: $f(\alpha)=q\alpha - \tau(q)$. The spectrum gives a quantitative measure of the amount of correlation present in a time series data. One can characterize the underlying processes of multifractality from a parametric representation of the singularity spectrum. Shimizu et al. \cite{Shimizu02} proposed a quadratic parametrization of $f(\alpha)$ around the position of maximum $\alpha_0$ such as
\beqn
f(\alpha) = A+B(\alpha-\alpha_0)+C(\alpha-\alpha_0)^2.
\label{quad}
\eeqn
Here $B$ is the asymmetry parameter, which is zero for symmetric, positive(negative) for left-(right-) skewed $f(\alpha)$ spectrum, and $A$ and $C$ are two parameters that govern the overall shape of the spectrum. Another parameter of interest is the width of the singularity spectrum $W = \alpha_{\max} -\alpha_{\min}$, can be obtained from a quadratic fit to the $f(\alpha)$ versus $\alpha$ data and then extrapolating it to $f(\alpha_{\max}) = f(\alpha_{\min}) = 0$. Parameters $\alpha_0$, $B$ and $W$ are used as a measure of ``complexity'' of the process under consideration \cite{Muoz06}, and hence they are called the ``complexity parameters''. Roughly specking, a smaller value of $\alpha_0$ implies the underlying process is more regular in appearance. The width $W$ measures the range of fractal exponents obtainable in the signal. So it gives the degree of multifractality of the signal. The wider the range of the fractal exponents (wider $f(\alpha)$ spectrum) corresponds to the richer structure of the process. The skewness in the shape of $f(\alpha)$ spectrum may be quantified by the ratio \cite{Muoz06}
\beqn
r = (\alpha_{\max} - \alpha_0)/(\alpha_0-\alpha_{\min}).
\eeqn
For a symmetric spectrum $r=1$, for right skewed $r>1$ and for left skewed $r<1$. The asymmetry parameter $B$ indicates which fractal exponents are dominant -- a right-skewed spectrum is dominated by high fractal exponents (process characterized by ``fine structure''), while a left skewed spectrum indicates the dominance of low fractal exponents (more regular or smooth looking process). In summary, a signal with a high value of $\alpha_0$, a wide range of fractal exponents (higher $W$) and a right-skewed $(B < 0)$, may be considered more complex than those with opposite characteristics \cite{Shimizu02}. However, a quadratic function may not always explain well the observed $f(\alpha)$ spectrum, as it is also observed here. In such cases, a fourth degree polynomial
\beqn
f(\alpha) = A + B(\alpha-\alpha_0) + C(\alpha-\alpha_0)^2 +D(\alpha-\alpha_0)^3+E(\alpha-\alpha_0)^4
\label{eq:4th}
\eeqn
is used \cite{Muoz06}. The asymmetry in this case depends on the first and third order coefficients, respectively $B$ and $D$.

\begin{figure}[t]
\centering
\includegraphics[width=6.5in,height=7in]{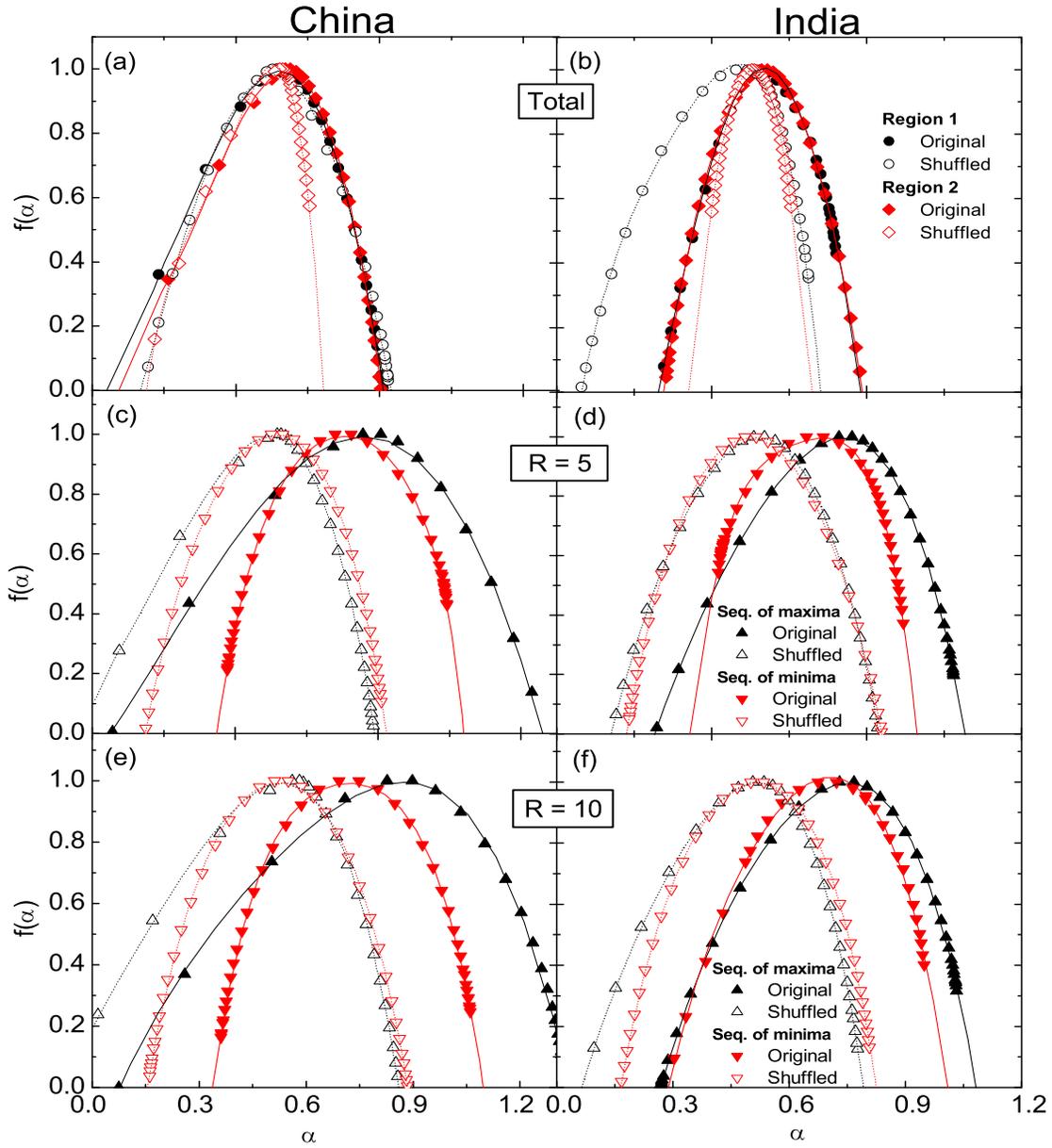}
\caption{The multifractal singularity spectra for the gold price returns in China and India, upper panel: the original series, middle panel: sequences of maximal (minimal) returns for $R=5$ and lower panel: sequences of maximal (minimal) returns for $R=10$. The lines represent the fourth-degree polynomial (\ref{eq:4th}) fitted to the data points. The complexity parameters obtained from the coefficients of the polynomial fits are given in Table \ref{Tab1}.} 
\label{Fig:Spec}
\end{figure}

The singularity spectra $f(\alpha)=q\alpha - \tau(q)$ calculated here are for the gold market return data as well as for their maxima (minima) sequences over consecutive non-overlapping windows of length $R=5$ and $R=10$. These spectra are shown in Fig. \ref{Fig:Spec} against the singularity strength $\alpha$, also known as the H\"older exponent. The predictions of the shuffled series/sequences corresponding to each of the empirical series/sequences are also included in the figure. One can see the $f(\alpha)$ spectra [Fig. \ref{Fig:Spec}(a)--(b)] corresponding to the original series are much narrower than those generated from the maxima (minima) of sequences [Fig. \ref{Fig:Spec}(c)--(f)]. Also exist very small difference between the original and the shuffled series generated $f(\alpha)$ spectra. The observation indicates that the original series possess almost as much multifractality originating from a fat-tailed probability distribution as any time series if it is generated at random. Conversely, the gold markets show no long-range persistence of memory. However, this feature is consistently found for the gold market returns. Actually, the markets contain a substantial amount of long-range correlation embedded into their volatility clustering.

In order to quantify of the complexity present in the gold price time series data, we fit the singularity spectra to the fourth-degree polynomial (\ref{eq:4th}) and use the fit parameters $A,\,B,\,\cdots$ etc. to extract the complexity parameters. Note that the quadratic function (\ref{quad}) is unable to replicate the spectra. The calculated values of the complexity parameters $\alpha_0$, $W$ and $r$, are given in Table \ref{Tab1}, where each parameter obtained from the real data is followed by its shuffled counterpart (shown within parenthesis). Results for the maxima (minima) sequence obtained on daily basis, and for $R=5\;\mbox{and}\; 10$ are shown in the table. It is seen that in most cases the $\alpha_0$ and $W$ values calculated from the real sequences for $R=5\;\mbox{and}\; 10$ are always greater than those calculated from the shuffled series, while all real series spectra are left-skewed ($r<1$). The shuffled series generated $f(\alpha)$ spectra, as expected, are all peaked at $\alpha_0 \approx 0.5$ and their widths ($W$) are also consistently less than the corresponding original series generated spectrum. For the skewness parameter $r$ we do not see any such systematic difference between the original and the shuffled series. Probably, this parameter itself is not very sensitive to market fluctuation data. Based upon the parameter $\alpha_0$ and $W$ we may argue that the complexity of the sequence of the maximal returns is slightly higher than that of the minimal returns, and that the overall complexity of the original time series is enhanced while subjected to $R=5\;\mbox{and}\; 10$ sequences. The sequences with $R=5$ and $10$ possess almost identical complexity. From Table \ref{Tab1} we also note that both for $R=5$ and 10 the $\alpha_0$ and $W$ values for the Chinese market are consistently higher than those of the Indian market, which indicates that the underlying dynamics of the former is a little more complex than that of the later.
\begin{table}
\caption{The complexity parameters $\alpha_0$, $W$ and $r$ for the analyzed gold market time series and their sequences (Seq.) of maxima (minima) for two different values of time window $R=5$ and 10. The original series/sequence estimated values follow their shuffled series/sequence estimates, given under parentheses.}
\begin{center}
\begin{tabular}{llll|lll}
    \hline
    & \multicolumn{3}{c}{China (CNY)}&\multicolumn{3}{c}{India (INR)}\\
    \hline
    Series/Sequence & $~\alpha_0$ & $~W$ & $~r$  &  $~\alpha_0$ & $~W$ & $~r$\\
    \hline
   Total (Region I): &~0.540 & ~0.709&~0.410 & ~0.530 & ~0.507 & ~0.920 \\
                    & (0.510) & (0.691) & (0.813) &(0.499) &(0.618)&(0.489)\\
   Total (Region II): &~0.535 &~0.720 &~0.583 &~0.550 &~0.510 &~0.867\\
               &(0.510) & (0.652)& (0.801)& (0.497)&(0.311)& (1.060)\\          
    \hline
Seq. maxima (R=5):&~0.760& ~1.215& ~0.701  &~0.742 &~0.800 &~0.613 \\
             &(0.510)&(0.829)&(0.516) &(0.500)& (0.681)& (0.907)\\
Seq. minima (R=5): &~0.709 &~0.690 &~0.912 &~0.680 &~0.578 &~0.751\\
             &(0.510)& (0.677)& (0.849) &(0.500)& (0.647)& (1.041)\\
     \hline
Seq. maxima (R=10):   &~0.850 &~1.266 &~0.613   &~0.756 &~0.823 &~0.663   \\
                      &(0.520)& (1.023)& (0.268)&(0.510)& (0.729)& (0.634)\\ 
Seq. minima (R=10):                        
                    &~0.720 &~0.756 &~0.994   &~0.708 &~0.717 &~0.701\\
                    &(0.520)& (0.728)& (0.982) &(0.510) &(0.654)& (0.895)\\
    \hline
\end{tabular}
\end{center}
\label{Tab1}
\end{table}

\begin{figure}[t]
\centering
\includegraphics[width=6.5in,height=5in]{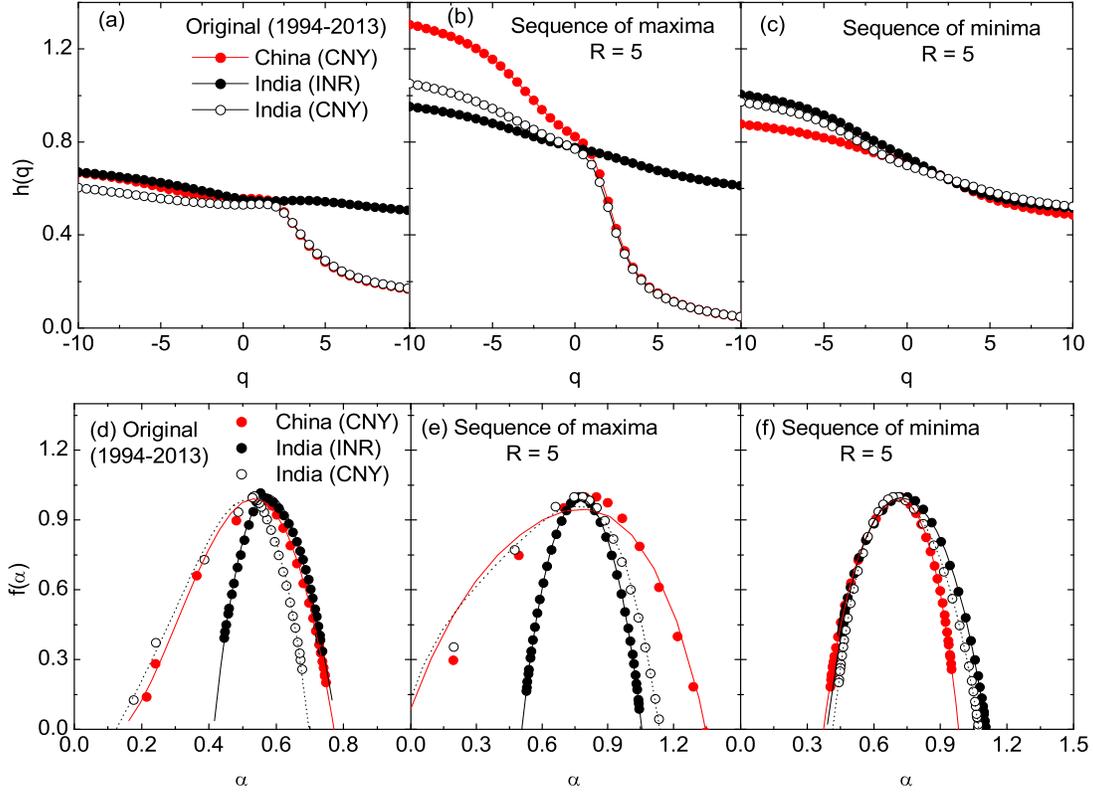}
\caption{Spectra of the multifractal variables $h(q)$ and $f(\alpha)$ for the period 1994-2013. Here the effect of currency conversion form INR to CNY is examined.} 
\label{Fig:New}
\end{figure}

As mentioned before, the relative variation of one local currency with respect to the other can influence the results. Therefore, we also investigate how the gold markets of China and/or India are affected by the currency fluctuations between the two countries. For this purpose we analyze the market returns over a period of 1994-2013 by converting the Indian Rupee series into a Chinese Yuan \cite{OANDA}. Prior to October 1993 the conversion factors are not available to us. However, even with a subset of the data it is still possible to qualitatively understand how the relative currency fluctuations affects the fluctuations of local gold markets. The results of this analysis is numerically presented in Table \ref{Tab2} and schematically in Fig. \ref{Fig:New}. It is found that the $h(q)$ and/or $f(\alpha)$ spectra for the Indian market series (black full circles), when calculated in terms of the Chinese currency (black empty circles), consistently shift towards the respective spectra for the Chinese series (red full circles). The relative differences between the multifractal parameters of these two markets are also reduced significantly. This is a manifestation of the (cross)correlative nature of two identical market series. Such correlations can also be studied in terms of the so-called cross-correlated MF-DFA analysis \cite{Horvatic11,Podobnik09}.  

\begin{table}
\caption{The complexity parameters $\alpha_0$, $W$ and $r$ for the gold market returns over the period 1994-2013. The effect of currency conversion on the complexity parameters is illustrated.}
\begin{center}
\begin{tabular}{llll|lll|lll}
    \hline
    & \multicolumn{3}{c}{China (CNY)}&\multicolumn{3}{c}{India (INR)} & \multicolumn{3}{c}{India (CNY)} \\
    \hline
    Series/Sequence & $~\alpha_0$ & $~W$ & $~r$  &  $~\alpha_0$ & $~W$ & $~r$ &  $~\alpha_0$ & $~W$ & $~r$ \\
    \hline
Original series & 0.532 & 0.602 &  0.627 & 0.548 & 0.372 & 1.747 & 0.531 & 0.575  & 0.428 \\
Seq. maxima  & 0.848 & 1.340 &  0.509 & 0.751 & 0.582 & 1.208 & 0.778 & 1.172 & 0.433 \\
Seq. minima  & 0.711 & 0.615 &  0.820 & 0.710 & 0.719 & 1.234 & 0.747 & 0.649 & 0.990 \\
\hline
\end{tabular}
\end{center}
\label{Tab2}
\end{table}
\section{Conclusions}
In this article we study the effects of long-term memory in the gold market time series of China and India over the period 1985-2013. We analyze the autocorrelation functions for these market series and find that the market series do not exhibit any (long- and/or short-range) correlation. However, the multifractal detrended fluctuation analysis (MF-DFA) method reveals a weak multifractal structure for those market returns and one seems that the major contribution to the observed underlying processes arises from the fat-tailed probability density function of the values. This is confirmed by a direct comparison between the multifractal results of the original and the corresponding shuffled series. Our observations in this regard contradict the results of the previous publications based on gold markets \cite{Bolgorian11,Wang11,Ghosh12}, where an existence of long-range correlation in the gold market data is claimed. In this analysis we show that the multifractality in the gold market time series for the period 1985-2013 is not because of the long-range correlation of the daily returns, but it is due to the correlation present in the sequences of maximal and minimal returns in successive non-overlapping time windows. The MF-DFA method is also applied to characterize the multifractal nature of these sequences.\\ 
\\
Our analysis on the sequences of maximal (minimal) returns shows enough indication of long-range correlation in the data. Multifractality in the sequences is found to be due to both the long-range persistence of memory and to the fat-tail probability density function, and the strength of multifractality in the sequences are much more stronger than what was obtained for the original series. We parametrize the corresponding multifractal patterns in terms of a set of complexity parameters \cite{Shimizu02,Muoz06} and find that the sequence series are more complex than the actual series from where the sequences are extracted. The complexity of the positive sequences is found to be higher than that of the negative sequences. Overall complexities do not alter appreciably with increasing window length from 5 to 10 trading days. The observation indicates that the gold market returns sustain a long-range memory, at least over a period of a few weeks. A significant component of the difference between the Chinese and the Indian gold market can be attributed to the relative fluctuations between the local currencies. However, it appears that the complexities of the Chinese gold market is more than those of the Indian market. Right now it is difficult to draw any definite conclusion about what are the exact implications of the results obtained from the present analysis on the gold trading policies of the two countries under consideration. Perhaps a more rigorous analysis performed over different time periods taking the economic policies adopted by the two countries into account will be more useful in this regard.
\section*{References}
\bibliographystyle{plain}

\end{document}